\documentclass[amsmath, amsfonts, showpacs, superscriptaddress, twocolumn, prb]{revtex4}
\usepackage{graphicx}
\usepackage{epsfig}
\usepackage{bm}
\usepackage{dcolumn}
\usepackage{amsmath}
\usepackage{amssymb}

\usepackage{color}
\usepackage{stackrel}
\usepackage{accents}

\usepackage{latexsym}

\usepackage{verbatim}

\usepackage{hyperref}

\newcommand{\bsub}{\begin{subequations}}
\newcommand{\esub}{\end{subequations}}

\newcommand \bea {\begin{eqnarray} }
\newcommand \eea {\end{eqnarray}}
 
\newcommand{\beg}{\begin{equation}}
\newcommand{\en}{\end{equation}}
\newcommand{\bp}{\mathbf p}
\newcommand{\bq}{\mathbf q}

\newcommand \bel  {\begin{align}}
\newcommand \enl  {\end{align}}

\newcommand{\up}{\uparrow}
\newcommand{\dn}{\downarrow}
\newcommand{\dg}{^\dagger}

\newcommand{\bk}{\mathbf k}

\newcommand{\pmat}{\begin{pmatrix}}
\newcommand{\epmat}{\end{pmatrix}}

\def\8{\infty}

\def\undertext#1{\vtop{\hbox{#1}\kern 1pt \hrule}}

\def\be{\begin{equation}}
\def\ee{\end{equation}}
\def\bea{\begin{eqnarray} & &}
\def\eea{\end{eqnarray}}

\begin{document}

\title{Amplitude modes and dynamic coexistence of competing orders \\ in multicomponent superconductors}

\author{Maxim Dzero}
\affiliation{Department of Physics, Kent State University, Kent, Ohio 44242, USA}

\author{Maxim Khodas}
\affiliation{Department of Physics and Astronomy, University of Iowa, Iowa City, Iowa 52242, USA}
\affiliation{Racah Institute of Physics, Hebrew University of Jerusalem, Jerusalem 91904, Israel}

\author{Alex Levchenko}
\affiliation{Department of Physics, University of Wisconsin-Madison, Madison, Wisconsin 53706, USA}
\affiliation{Department of Physics and Astronomy, Michigan State University, East Lansing, Michigan 48824, USA}

\begin{abstract}
We study the nonequilibrium dynamics of an electronic model with competing spin-density-wave and unconventional superconductivity in the context of iron pnictides. Focusing on the collisionless regime, we find that magnetic and superconducting order parameters may coexist dynamically after a sudden quench, even though the equilibrium thermodynamic state supports only one order parameter. We consider various initial conditions concomitant with the phase diagram and in a certain regime identify different oscillatory amplitude modes with incommensurate frequencies for magnetic and superconducting responses. At the technical level we solve the equations of motion for the electronic Green's functions and self-consistency conditions by reducing the problem to a closed set of Bloch equations in a pseudospin representation. For certain quench scenarios the nonadiabatic dynamics of the pairing amplitude is completely integrable and in principle can be found exactly.   
\end{abstract}

\date{June 2, 2015}


\maketitle

\section{Introduction}

Conventional superfluids and superconductors host various collective oscillations. The best studied examples include the phase mode of the order parameter (OP), the so-called Anderson-Bogoliubov mode,~\cite{Anderson,Bogolubov} and the amplitude oscillations in the magnitude of the superconducting gap, the so-called longitudinal Schmid mode.~\cite{Schmid,Volkov-Kogan} In charged superfluids the coupled oscillations in the phase of the order parameter and the electric field appear because of gauge invariance. Physically, this mode corresponds to the balanced oscillations between the normal current and supercurrent, and in the literature it is called the transverse Carlson-Goldman mode.~\cite{Carlson-Goldman,Schmid-Schon} Early works on superconducting modes were comprehensively summarized by Artemenko and Volkov,~\cite{Artemenko} and Kulik, Entin-Wohlmant, and Orbach,~\cite{Kulik} including studies of disorder scattering effects and charge imbalance on the dispersion and attenuation of collective oscillations. 
 
In multicomponent systems or superconductors with unconventional symmetry of the OP, the plethora of collective effects is even richer.~\cite{Wolfle,Sauls,Balatsky,Higashitani,Ohashi,Sharapov} In multiband superconductors such as MgB$_2$ the oscillations of the phase difference of OPs between the two bands is charge neutral, in contrast to phase average plasma oscillations. A phase difference low frequency Leggett mode~\cite{Leggett} is an in-gap weakly damped excitation observed in the Raman response of MgB$_2$.~\cite{Blumberg-MgB2} It is natural to look for Leggett-like and Carlson-Goldman modes in multiband and iron-pnictide superconductors (FeSCs).~\cite{Burnell,Anishchanka} Normally, these modes are overdamped with frequencies well within the quasiparticle continuum. There are, however, important exceptions to this rule. A typical setting for this scenario is the change in the OP symmetry controlled by external parameters. In many cases such a transformation proceeds via an intermediate phase with broken time reversal symmetry (TRS).~\cite{Lee,Stanev-1,Babaev,Lin,Platt,Stanev-2,Maiti,Khodas,Marciani} Soft Leggett-like modes are found at the boundaries of the intermediate lower symmetry phase. A transformation of this kind was very recently induced by pressure in KFe$_2$As$_2$,~\cite{Tafti} and TRS breaking along with Leggett-like modes await experimental detection.

Different kinds of collective excitations are the Bardasis-Schrieffer modes.~\cite{Bardasis-Schrieffer,Vaks} These modes are carried by Cooper pairs accelerated to higher angular momentum states and manifest as in-gap excitons. Since all but the $s$-wave channels are charge neutral, Bardasis-Schrieffer modes remain low energy in-gap excitations even in the presence of Coulomb repulsion. As photons transfer the angular momentum to Cooper pairs, Raman spectroscopy~\cite{Devereaux,Klein} is ideally suited to probe Bardasis-Schrieffer modes.~\cite{Klein-Dierker,Monien,Chubukov,Scalapino,Kretzschmar} 

Interestingly, Raman spectroscopy was originally suggested as a tool to detect amplitude Higgs modes, whereas they were indirectly observed owing to coupling to intermediary collective excitations.~\cite{Sooryakumar,Littlewood-1,Littlewood-2} More recent Raman,~\cite{Measson} terahertz pump-probe spectroscopy,~\cite{Matsunaga} and combined tunneling and optical conductivity measurements~\cite{Frydman} provide unambiguous direct tests of massive Higgs modes in superconductors. In a parallel vein, the coherent amplitude mode has been observed in the strongly interacting superfluid phases of bosonic ultracold atoms in optical lattices by means of Bragg spectroscopy and lattice modulation.~\cite{Bissbort,Bloch} All these findings have stimulated many theoretical efforts, (see a recent review article in Ref.~[\onlinecite{Pekker}] and references therein).     

Since the pioneering work by Volkov and Kogan,~\cite{Volkov-Kogan} persistent oscillations of the superconducting OP have been predicted to appear in a response to a fast nonadiabatic perturbation (quench) in the collisionless regime.~\cite{BLS,BL,Dzero,Tsyplyatyev,Gurarie,YDGF} The mode frequency is determined by the superconducting gap whereas oscillations are superimposed with a slow power-law decay. In contrast, theoretical studies of nonequilibrium dynamics after ultrafast excitation in complex superconducting systems hosting coexisting OPs are in their early stages, with only a few recent results.~\cite{Eremin,Moor,Sachdev}  The main thrust of this paper is to provide a detailed description of the coupled dynamics of amplitude modes in the context of FeSC systems. Broadly formulated, our theory may shed  light on the hotly debated issue of the structure of the OP and the closely related question of the competition between magnetism and superconductivity in FeSCs as seen out of equilibrium. Our motivation comes from  recent ultrafast measurements that reveal a dynamic coexistence and interplay of multiple order parameters in various strongly correlated materials.~\cite{Orenstein-1,Orenstein-2,Torchinsky,Patz,Pogrebna}

This paper is organized as follows. In Sec.~\ref{Sec-Model} we adopt the simplest model of iron pnictides, where the dynamics of competing orders is already found to display a nontrivial character. We briefly discuss the ground state properties of this model and derive the equations of motion for the Green's functions in the pseudospin representation. In Sec.~\ref{Sec-Dynamics} we numerically integrate these coupled equations together with self-consistency constraints, and discuss the emergent dynamic coexistence of superconductivity and magnetism. Section~\ref{Sec-Lax} is devoted to the analysis of a special case when the dynamics of the order parameters is integrable. We summarize our results in Sec.~\ref{Sec-Discussion} and place our work into the perspective of future studies.

\section{Model}\label{Sec-Model}
To study th- dynamical interplay between superconductivity and spin-density wave order, we use the minimal two-band model previously introduced in the context of iron-pnictide superconductors. Following  Refs.~[\onlinecite{VVC2010,FS2010}], we consider the Hamiltonian 
\beg\label{H}
\hat{H}=\hat{H}_0+\hat{H}_\Delta+\hat{H}_m.
\en
The first term accounts for the electronlike and holelike two-dimensional (2D) fermionic bands,
\beg\label{H0}
\hat{H}_0=\sum\limits_{\bk}\left\{\xi_{\bk c}{c}_{\bk\alpha}\dg c_{\bk\alpha}+\xi_{\bk'f}{f}_{\bk'\alpha}\dg
f_{\bk'\alpha}\right\},
\en
where $f_{\bk'\alpha}^\dagger$, $f_{\bk'\alpha}$ (${\bk'}=\bk-{\mathbf Q}$) are the creation and annihilation operators for the fermions with a spin projection $\alpha$ near an electron pocket ${\mathbf Q}=(0,\pi)$ of the two-dimensional Brillouin zone with dispersion 
\beg\label{xif}
\xi_{\bk f}=\frac{k^2}{2}-\mu_f,
\en
the chemical potential $\mu_f$, and we set the electron's mass to one. 
Similarly, $c_{\bk\alpha}^\dagger$, $c_{\bk\alpha}$ describe the fermions near the $\Gamma=(0,0)$ point with the hole band with a chemical potential $\mu_c$ and dispersion 
\beg\label{xic}
\xi_{\bk c}=\mu_c-\frac{k^2}{2}.
\en
The second term in Eq.~(\ref{H}) accounts for the superconducting pairing. Within the mean-field approximation we have
\beg\label{HDelta}
\begin{split}
\hat{H}_\Delta=&\frac{1}{2}\sum\limits_\bk\left\{\Delta_{\alpha\beta}^c{c}_{\bk\alpha}\dg{c}_{-\bk\beta}\dg+
\Delta_{\alpha\beta}^f{f}_{\bk\alpha}\dg{f}_{-\bk\beta}\dg+\textrm{h.c.}
\right\}, 
\end{split}
\en
where $\Delta_{\alpha\beta}^{c,f}$ are the superconducting order parameters defined for each band,
\beg\label{DLTfDLTc}
\begin{split}
&\Delta_{\alpha\beta}^c=g_{sc}\sum\limits_{\bk}(i\sigma^y)_{\alpha\beta}(i\sigma^y)_{\gamma\delta}\dg
\langle f_{-\bk\gamma}f_{\bk\delta}\rangle, \\ 
&\Delta_{\alpha\beta}^f=g_{sc}\sum\limits_{\bk}(i\sigma^y)_{\alpha\beta}(i\sigma^y)_{\gamma\delta}\dg\langle c_{-\bk\gamma}c_{\bk\delta}\rangle,
\end{split}
\en
and $g_{sc}>0$ is the superconducting coupling. Finally, the last term in Eq.~(\ref{H}) describes the onset of the commensurate spin-density-wave (SDW) order, which within the mean-field approximation is described by 
\beg\label{Hm}
\hat{H}_m=\frac{1}{2}\sum\limits_\bk m_{\alpha\beta}\left\{{f}_{\bk\alpha}\dg c_{\bk\beta}+
{c}_{\bk\alpha}\dg f_{\bk\beta}\right\}+\textrm{h.c.},
\en
where the SDW order parameter is determined self-consistently via
\beg\label{selfM}
m_{\alpha\beta}=-\frac{g_{m}}{2}\sum\limits_\bp{\vec \sigma}_{\alpha\beta}\cdot{\vec \sigma}_{\gamma\delta}\dg
\langle{c}_{\bp\gamma}\dg f_{\bp\delta}\rangle,
\en
and $g_{m}>0$ is the corresponding coupling constant. In what follows, without loss of generality, we assume that (i) there is no mismatch between the hole and electron Fermi surface, $\mu_c=\mu_f$, (ii) $m_{\alpha\beta}=m_z\sigma_{\alpha\beta}^z$, and (iii) the superconductivity is of $s^{\pm}$ type, so that $\Delta_{\alpha\beta}^{c,f}=\Delta_{c,f}(i\sigma^y)_{\alpha\beta}$ and
\beg\label{spm}
\Delta_c=-\Delta_f=\Delta.
\en
Below, we briefly review the ground state properties of the mean-field model (\ref{H}) for the $s^{\pm}$ superconducting pairing.

\subsection{Correlation functions and pseudospins}
In order to obtain the equations of motion, which determine the nonadiabatic dynamics for superconducting and magnetic order parameters, we introduce the four-component spinor 
\beg\label{Psi}
{\Psi}_{\bk\alpha}\dg=\left(~{c}_{\bk\alpha}\dg, ~c_{-\bk\alpha},~{f}_{\bk\alpha}\dg, ~f_{-\bk\alpha}~\right).
\en
The corresponding real-time correlation functions are 
\beg\label{Gab}
\hat{G}_{\alpha\beta}(\bk;t_1,t_2)=-i\left\langle\hat{T}\left\{\Psi_{\bk\alpha}(t_1)\Psi_{\bk\beta}\dg(t_2)\right\}\right\rangle.
\en
We further consider normal-$G$ and anomalous-$F$ propagators for $c$-fermions, 
\beg\label{normanomc}
\begin{split}
&G_{\alpha\beta}^{c}(\bk;t_1,t_2)=-i\langle\hat{T}\{c_{\bk\alpha}(t_1){c}_{\bk\beta}\dg(t_2)\}\rangle, \\
&F_{\alpha\beta}^{c}(\bk;t_1,t_2)=-i\langle\hat{T}\{c_{\bk\alpha}(t_1){c}_{-\bk\beta}(t_2)\}\rangle, \\
&\overline{F}_{\alpha\beta}^{c}(\bk;t_1,t_2)=-i\langle\hat{T}\{{c}_{-\bk\alpha}\dg(t_1){c}_{\bk\beta}\dg(t_2)\}\rangle,\\
&\widetilde{G}_{\alpha\beta}^c(\bk;t_1,t_2)=-i\langle\hat{T}\{{c}_{-\bk\alpha}\dg(t_1){c}_{-\bk\beta}(t_2)\}\rangle,
\end{split}
\en
and analogously for $f$-fermions. In addition, we also consider the mixed correlators
\beg\label{mixed}
\begin{split}
&G_{\alpha\beta}^{m}(\bk,\bq;t_1,t_2)=-i\langle\hat{T}\{c_{\bk\alpha}(t_1){f}_{\bk\beta}\dg(t_2)\}\rangle, \\
& \widetilde{G}_{\alpha\beta}^{m}(\bk;t_1,t_2)=-i\langle\hat{T}\{f_{\bk\alpha}(t_1){c}_{\bk\beta}\dg(t_2)\}\rangle.
\end{split}
\en

The correlation functions above depend on $t_1$ and $t_2$. However, the magnetic and superconducting order parameters are determined at $t_1=t_2$ and therefore will depend on $t=(t_1+t_2)/2$ only. 
Accordingly, we introduce the following pseudospin variables:~\cite{Anderson}
\beg\label{spins}
\begin{split}
&K_c^{-}(\bk,t)=K_{c}^x(\bk,t)-iK_c^y(\bk,t)=iF_{\dn\up}^c(\bk,t),  \\ 
&K_f^{-}(\bk,t)=K_{f}^x(\bk,t)-iK_f^y(\bk,t)=iF_{\dn\up}^f(\bk,t), \\ 
& K_f^{+}(\bk,t)=i\overline{F}_{\up\dn}^{f}(\bk,t), \quad K_c^{+}(\bk,t)=i\overline{F}_{\up\dn}^{c}(\bk,t),\\
&K_{c,f}^z(\bk,t)=-\frac{i}{2}\sum\limits_{\alpha=\up,\dn}G_{\alpha\alpha}^{c,f}(\bk,t).
\end{split}
\en
Similarly, we introduce the additional pseudospins ${\vec S}(\bk,t)$ and ${\vec N}(\mathbf{k},t)$, which are defined by the mixed averages from Eq.~(\ref{mixed}):
\beg\label{SpinsSN}
\begin{split}
&S_x(\bk,t)=\frac{i}{2}\sum\limits_{\alpha=\up,\dn}\left\{{G}_{\alpha\overline{\alpha}}^{m}(\bk;t)+\widetilde{G}_{\alpha\overline{\alpha}}^{m}(\bk;t)\right\}, \\ &S_y(\bk,t)=-\frac{1}{2}\sum\limits_{\alpha=\up,\dn}\textrm{sign}(\alpha)\left\{{G}_{\alpha\overline{\alpha}}^{m}(\bk;t)+\widetilde{G}_{\alpha\overline{\alpha}}^{m}(\bk;t)\right\}, \\
&S_z(\bk,t)=\frac{i}{2}\sum\limits_{\alpha=\up,\dn}\textrm{sign}(\alpha)\left\{{G}_{\alpha\alpha}^{m}(\bk;t)+\widetilde{G}_{\alpha{\alpha}}^{m}(\bk;t)\right\},\\
&N_x(\bk,t)=\frac{1}{2}\sum\limits_{\alpha=\up,\dn}\left\{{G}_{\alpha\overline{\alpha}}^{m}(\bk;t)-\widetilde{G}_{\alpha\overline{\alpha}}^{m}(\bk;t)\right\}, \\ &N_y(\bk,t)=\frac{i}{2}\sum\limits_{\alpha=\up,\dn}\textrm{sign}(\alpha)\left\{{G}_{\alpha\overline{\alpha}}^{m}(\bk;t)-\widetilde{G}_{\alpha\overline{\alpha}}^{m}(\bk;t)\right\}, \\
&N_z(\bk,t)=\frac{1}{2}\sum\limits_{\alpha=\up,\dn}\textrm{sign}(\alpha)\left\{{G}_{\alpha\alpha}^{m}(\bk;t)-\widetilde{G}_{\alpha{\alpha}}^{m}(\bk;t)\right\}.
\end{split}
\en
Finally, we will also need pseudospins ${\vec L}(\bk,t)$, which are defined according to
\beg
\begin{split}
&L_x(\bk,t)=-\frac{i}{2}\sum\limits_{\alpha=\up,\dn}\left\{{G}_{\alpha\overline{\alpha}}^{c}(\bk;t)+{G}_{\alpha\overline{\alpha}}^{f}(\bk;t)\right\}, \\ 
&L_y(\bk,t)=\frac{1}{2}\sum\limits_{\alpha=\up,\dn}\textrm{sign}(\alpha)\left\{{G}_{\alpha\overline{\alpha}}^{c}(\bk;t)+\widetilde{G}_{\alpha\overline{\alpha}}^{f}(\bk;t)\right\}, \\
&L_z(\bk,t)=-\frac{i}{2}\sum\limits_{\alpha=\up,\dn}\textrm{sign}(\alpha)\left\{{G}_{\alpha\alpha}^{c}(\bk;t)+{G}_{\alpha\alpha}^{f}(\bk;t)\right\}.
\end{split}
\en

Equations of motion for the pseudospins can be obtained from the equations of motion for the fermionic operators. Using the Heisenberg representation
\beg
c_{\bk\alpha}(t)=e^{i\hat{H}t}c_{\bk\alpha}e^{-i\hat{H}t},
\en
we have
\beg
\begin{split}
&i\frac{\partial}{\partial t}c_{\bk\alpha}=\xi_c(\bk)c_{\bk\alpha}+
\Delta_{\alpha\overline{\alpha}}^c{c}_{-\bk\overline{\alpha}}\dg+\sum\limits_\beta m_{\alpha\beta}f_{\bk\beta}, \\
&i\frac{\partial}{\partial t}{c}_{\bk\beta}\dg=-\xi_c(\bk){c}_{\bk\beta}\dg-
c_{-\bk\overline{\beta}}\overline{\Delta}_{\overline{\beta}\beta}^c -
\sum\limits_\alpha {f}_{\bk\alpha}\dg m_{\alpha\beta},
\end{split}
\en
where we used $\xi_c(\bk)=\xi_c(-\bk)$. Similarly, the equations for the $f$-operators are
\beg
\begin{split}
&i\frac{\partial}{\partial t}f_{\bk\alpha}=\xi_f(\bk)f_{\bk\alpha}+
\Delta_{\alpha\overline{\alpha}}^f{f}_{-\bk\overline{\alpha}}\dg+
\sum\limits_\beta m_{\alpha\beta}c_{\bk\beta}, \\
&i\frac{\partial}{\partial t}{f}_{\bk\beta}\dg=-\xi_f(\bk){f}_{\bk\beta}\dg-
\overline{\Delta}_{\overline{\beta}\beta}^f f_{-\bk\overline{\beta}}-
\sum\limits_\alpha {c}_{\bk\alpha}\dg m_{\alpha\beta}.
\end{split}
\en
Here $\overline{\Delta}$ denotes the complex conjugate of $\Delta$ and $\overline{\alpha}=-\alpha$. 
From these equations we derive the equations of motion for the correlators above. 

\subsection{Ground state}
In this section we discuss the ground state properties of the mean-field Hamiltonian (\ref{H}). Generally, the ground state properties can be derived by analyzing the free energy within the Luttinger-Ward's generating functional method.~\cite{Luttinger1960} For convenience we adopt the pseudospin variables. The distribution of the pseudospins variables ${\vec K}_{\bk c,f}$ with respect to momentum follows from the mean-field theory. For example, for the $K_{\bk f,c}^z$ we find 
\beg\label{Kz}
\begin{split}
K_{\bk f,c}^z&=\frac{\xi_{\bk c,f}m_z^2-\xi_{\bk f,c}\left(\xi_{\bk c,f}^2+\Delta^2+|E_{\bk+}E_{\bk-}|\right)}{2\left(E_{\bk+}^2|E_{\bk-}|+E_{\bk-}^2|E_{\bk+}|\right)}.
\end{split}
\en
Here we have introduced the renormalized quasiparticle spectrum
\beg\label{Epmk}
E_{\bk\pm}=\sqrt{\xi_\bk^2+\delta_\bk^2+m_z^2+\Delta^2\pm 2|\delta_\bk|\sqrt{m_z^2+\xi_\bk^2}}
\en
with $\xi_\bk=(\xi_{\bk f}-\xi_{\bk c})/2$ and function $\delta_\bk=(\xi_{\bk f}+\xi_{\bk c})/2$ which accounts for the Fermi surface mismatch. The anomalous $x$-components of the ${\vec K}$ are given by
\beg\label{Kxy}
\begin{split}
K_{\bk f,c}^x&=\frac{\Delta\left(|E_{\bk +}E_{\bk -}|+\Delta^2+\xi_{\bk c,f}^2-m_z^2\right)}{2\left(E_{\bk+}^2|E_{\bk-}|+E_{\bk-}^2|E_{\bk+}|\right)},
\end{split} 
\en
and the remaining components are zero. Setting in the equations above $\delta_\bk=0$, the pairing amplitude is determined by the corresponding self-consistency condition
\beg\label{SelfConsisten}
\Delta=-g_{sc}\sum\limits_{\bp}K_{\bp c}^{-}=g_{sc}\sum\limits_{\bp}K_{\bp f}^{-},
\en
where $K_{\bp a}^{-}=K_{\bp a}^x-iK_{\bp a}^y$. 

Similarly, pseudospin variables ${\vec S}_\bk$ are determined by the mixed correlators. In the ground state $S_{\bk}^{x,y}=0$ and
\beg\label{initS}
S_{\bk}^z=\frac{m_z\left(|E_{\bk+}E_{\bk-}|+m_z^2-\xi_{\bk f}\xi_{\bk c}+\Delta^2\right)}{E_{\bk+}^2|E_{\bk -}|+E_{\bk-}^2|E_{\bk+}|}.
\en
The SDW order parameter is determined by 
\beg\label{sceq4mz}
m_z=g_{m}\sum\limits_{\bk}S_{\bk}^z.
\en
Numerical analysis of the self-consistency equations above shows that for the zero Fermi surface mismatch the ground state is determined by the ratio of the corresponding coupling constants for the magnetic and superconducting orders. When the critical temperature of the SDW transition is higher than the superconducting critical temperature, $T_{m}>T_{c}$, the system orders magnetically, $m_z\not =0$, $\Delta=0$. If one allows for a finite Fermi surface mismatch, then there is a critical value for $\delta_\bk$ when the superconducting order becomes energetically favorable. Furthermore, the thermodynamic phase diagram contains an intermediate region where two order parameters coexist.~\cite{VVC2010,FS2010}

\subsection{Equations of motion}
Pseudospin variables happen to be very convenient to describe the nonequilibrium dynamics of the magnetic and superconducting order parameters. The same technique has been recently employed to describe the Higgs mode in conventional superconductors.~\cite{Tsuji-Aoki} We find that the dynamics can be fully accounted for by the five pseudospins, which have three components each. Pseudospins ${\vec K}_{c,f}(\bk,t)$ describe the dynamics of the electronic degrees of freedom of hole and electron bands respectively. The corresponding equations of motion are: 
\beg\label{EOMKcf}
\begin{split}
&\partial_t{\vec K}_{\bp c}={\vec B}_{\bp c}(t)\times{\vec K}_{\bp c}(t)+{\vec e}_z\left({\vec m}(t)\cdot{\vec N}_{\bp}(t)\right), \\
&\partial_t{\vec K}_{\bp f}={\vec B}_{\bp f}(t)\times{\vec K}_{\bp f}(t)-{\vec e}_z\left({\vec m}(t)\cdot{\vec N}_\bp(t)\right),
\end{split}
\en
where we introduced the effective magnetic fields
\beg
{\vec B}_{\bp a}=2(-\Delta_a^x(t),-\Delta_a^y(t),\xi_{\bp a}),
\en
and ${\vec N}_\bk(t)$ accounts for the influence of the magnetic ordering on the superconducting dynamics. The remaining equations of motion are
\beg\label{SNL}
\begin{split}
&\partial_t{\vec S}_\bp+2\xi_\bp{\vec N}_\bp(t)+2{\vec m}(t)\times{\vec L}_\bp(t)=0, \\
&\partial_t{\vec L}_{\bp}+2{\vec m}(t)\times{\vec S}_\bp(t)=0, \\
&\partial_t{\vec N}_\bp-2\xi_\bp{\vec S}_\bp(t)+2{\vec m}(t)\left[K_{\bp c}^z(t)-K_{\bp f}^z(t)\right]=0. \\
\end{split}
\en
Equations~(\ref{EOMKcf}) -- (\ref{SNL}) represent the main result of this section. We will analyze these equations numerically in Sec. III. In Sec. IV we consider the special case when the first two equations (\ref{EOMKcf}) decouple from the rest: this situation corresponds to quenches of the magnetic coupling constant to zero. In this scenario, the equations of motion can be integrated exactly and we prove the integrability of this particular case.

\begin{figure}
\includegraphics[scale=0.26,angle=0]{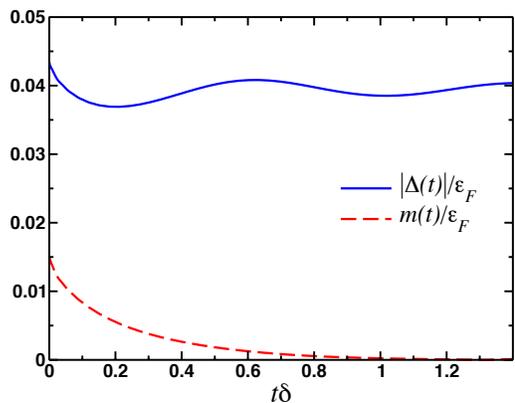}
\caption{(color online) Superconducting and magnetic order parameter dynamics for the metastable initial conditions for the choice of coupling constants corresponding to the ratio of critical temperatures $T_c=1.75T_{m}$. Both $\Delta(t)$ and $m(t)$ are given in units of the Fermi energy $\varepsilon_F$. We consider $N=1004$ pseudospins. The bandwidth $\varepsilon_\Lambda=10\varepsilon_F$ and the level spacing is $\delta=\varepsilon_\Lambda/N$.}
\label{Fig3}
\end{figure}

\section{Dynamical coexistence of superconductivity and SDW order}\label{Sec-Dynamics}

In this section we solve the equations of motion (\ref{EOMKcf}) and (\ref{SNL}) numerically. We consider the initial conditions corresponding to the metastable state of coexisting magnetism and superconductivity, $m_z\not=0$ and $\Delta\not=0$. For the ${\vec K}$ and ${\vec S}$ we choose the initial configuration corresponding to the metastable state where both $m_z$ and $\Delta$ are nonzero, Eqs. (\ref{Kz}), (\ref{Kxy}), and (\ref{initS}). In addition, for the initial conditions ${\vec m}=m_z{\vec e}_z$, we find ${\vec L}_\bk(t=0)={\vec N}_\bk(t=0)=0$. 

\begin{figure}
\includegraphics[scale=0.26,angle=0]{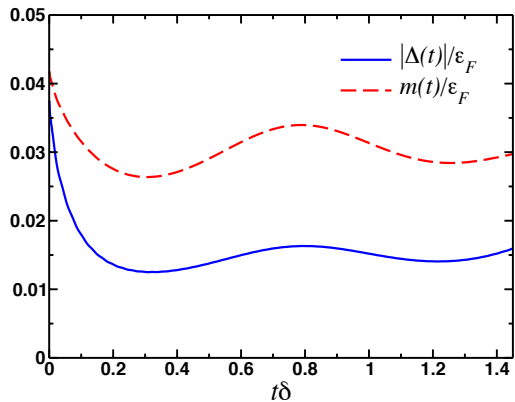}
\caption{(color online) Same as Fig. \ref{Fig3} for $T_c=0.95T_{m}$ }
\label{Fig4}
\end{figure}

\begin{figure}
\includegraphics[scale=0.26,angle=0]{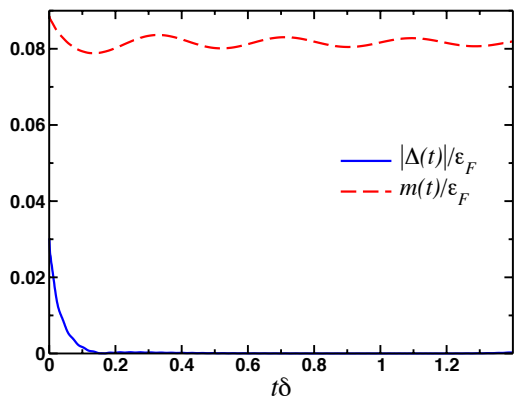}
\caption{(color online) Same as Fig. \ref{Fig3} for $T_c=0.45T_{m}$.}
\label{Fig5}
\end{figure}

We present the results of the numerical integration of the equations of motion in Figs.~\ref{Fig3}--\ref{Fig5}. In Fig.~\ref{Fig3} we choose the parameters corresponding to the superconducting ground state, $T_c=1.75T_{m}$. We see that in this case magnetization vanishes dynamically, while the pairing amplitude remains finite. 

The results in Fig.~\ref{Fig4} were obtained for $T_c=0.95T_{m}$. In this case we find that both magnetization and the pairing amplitude coexist dynamically. We observe that this nonequilibrium effect persists for the range of parameters corresponding to $T_c\simeq T_{m}$. 

Finally, in Fig.~\ref{Fig5} we show the time evolution of the pairing amplitude and magnetization when the initial values of magnetization and the superconducting energy gap are such that $T_c=0.5T_{m}$. In this case, we see that the pairing amplitude vanishes dynamically, while magnetization remains finite. 

The first and third scenarios are similar to previously studied cases of collisionless relaxation in a single-component system. The surviving order parameter, which corresponds to a thermodynamically favorable state, exhibits oscillatory behavior superimposed with a rather slow power-law decay at long times. The physical mechanism of relaxation is analogous to collisionless Landau damping in plasmas. The difference, however, is that in a gapped system such relaxation is typically nonexponential because of the branching singularity in the density of states. For example, the superconducting response was shown to fall asymptotically as $\propto \cos(2\Delta t)/\sqrt{t}$. 

The second scenario in Fig.~\ref{Fig4} is special and representative of the case when both order parameters are of comparable strength. Then the initial thermodynamically metastable state survives out of equilibrium for extended times until the system enters into the collision-dominated regime of relaxation. Depending on the choice of parameters in the model, both order parameters may oscillate with incommensurate frequencies. 

The dynamical effect of coexistence has been recently pointed out in the case of multiband superconductors, which can be applicable to either MgB$_2$ or iron-based superconductors in the part of the phase diagram without magnetism.~\cite{Eremin} In this case mutual dynamics is primarily triggered by the Josephson coupling of pair amplitudes between the bands. This is quite different as compared to the model which hosts order parameters whose physical nature is not the same. Known examples of the latter kind include the dynamical coexistence of bond-density-wave and $d$-wave superconductivity,~\cite{Sachdev} and charge-density-wave and $s$-wave superconductivity.~\cite{Moor} Our results expand these examples to the case of magnetically ordered systems. 

\section{Exactly solvable limit}\label{Sec-Lax}

In this section we consider quenches for which the nonadiabatic dynamics of the $s^{\pm}$ pairing amplitude can be found exactly. Specifically, we consider quenches into a state with a zero SDW order parameter, $m_z=0$. Formally, this limit can be realized for the quenches of the SDW coupling constant $g_{m}\to0$. 

The equations of motion are governed by the following Hamiltonian,
\beg\label{Hm0}
{\cal H}=2\sum\limits_{\bp i}\xi_{\bp,i}K_{\bp i}^z-\sum\limits_{\bp i}\left(\overline{\Delta}_i K_{\bp i}^{-}+\Delta_i K_{\bp i}^{+}\right),
\en
which we write in terms of the Anderson spins, where $i=c,f$. 

\subsection{Integrability criterion}

In order to demonstrate the exact integrability of the model (\ref{Hm0}), we adopt the method developed by Yuzbashyan \textit{et al.}~[\onlinecite{Enolskii-1,Enolskii-2}]. The central role in finding the dynamics of the pairing amplitude is played by the Lax vector. In order to identify the expression for the Lax vector for our problem, we first introduce
\beg\label{alam}
a_i=\left\{\begin{matrix} +1, \quad i=c \\ -1, \quad i=f
\end{matrix}
\right.
\en
From Eq. (\ref{spm}), it follows $\Delta_i=a_{\overline{i}} \Delta$. Using (\ref{alam}) we rewrite (\ref{SelfConsisten}) as 
\beg
2\Delta=g_{sc}\sum\limits_{\bp i}a_{\overline{i}} K_{\bp i}^-\equiv g_{sc}
\sum\limits_{\bp i} K_{\bp i}^-,
\en
where we used $a_{\overline{i}}=-a_i$, and redefined the anomalous components of the pseudospins
according to $a_{\overline{i}} K_{\bp i}^{x,y}\to K_{\bp i}^{x,y}$.
Clearly, this transformation leaves the Poisson brackets invariant:
\beg\label{PoissonBracket}
\{K_{\bp i}^a,K_{\bq j}^b\}=-\delta_{\bp,\bq}\delta_{ij}\epsilon^{abc}K_{\bp i}^c.
\en
Thus, the Hamiltonian (\ref{Hm0}) can now be rewritten as follows
\beg\label{Hm02}
{\cal H}=2\sum\limits_{\bp i}\xi_{\bp i}K_{\bp i}^z-\overline{\Delta}\sum\limits_{\bp i}K_{\bp i}^{-}-\Delta \sum\limits_{\bp i}K_{\bp i}^{+}.
\en

Consider the following Lax vector
\beg\label{Laxfc}
{\cal {\vec L}}(u)=\sum\limits_{\bp j}\frac{{\vec K}_{\bp j}}{u-\xi_{\bp j}}-\frac{2{\vec e}_z}{{g}_{sc}}.
\en
The Poisson brackets for the components of ${\vec {\cal L}}$ are obtained using (\ref{PoissonBracket}):
\beg\label{LaxPoisson}
\{{\cal L}^a(u),{\cal L}^b(v)\}=\epsilon^{abc}\frac{{\cal L}^c(u)-{\cal L}^c(v)}{u-v}.
\en
Due to the commutation relations ({\ref{LaxPoisson}) it follows that
\beg
\{{\vec {\cal L}}^2(u),{\vec {\cal L}}^2(v)\}=0.
\en
This property means that any model Hamiltonian which Poisson commutes with ${\cal L}^2$ will define an exactly integrable model.~\cite{Enolskii-1,Enolskii-2} Indeed, for the square of the Lax vector we readily find
\beg\label{L2}
\begin{split}
{\vec {\cal L}}^2(u)&=\sum\limits_{\bp i}\sum\limits_{\bq j}\frac{{\vec K}_{\bp i}\cdot{\vec K}_{\bq j}}{(u-\xi_{\bp i})(u-\xi_{\bq j})}+\frac{4}{g_{sc}^2}\\&-\frac{4}{g_{sc}}\sum\limits_{\bq i}\frac{K_{\bp i}^z}{u-\xi_{\bp i}}.
\end{split}
\en
The first term and the last terms should be rewritten as follows,
\beg\label{FirstTerm}
\begin{split}
&\sum\limits_{\bp i}\sum\limits_{\bq j}\frac{{\vec K}_{\bp i}\cdot{\vec K}_{\bq j}}{(u-\xi_{\bp i})(u-\xi_{\bq j})}-\frac{4}{g_{sc}}\sum\limits_{\bq i}\frac{K_{\bp i}^z}{u-\xi_{\bp i}}\\&=2\sum\limits_{\bp i}\frac{{\cal H}_{\bp i}}{u-\xi_{\bp i}}+\sum\limits_{\bp i}\frac{{\vec K}_{\bp i}^2}{(u-\xi_{\bp i})^2},
\end{split}
\en
where
\beg\label{calHp}
{\cal H}_{\bp i}=\sum\limits_{\bq\not=\bp}\sum\limits_{j\not= i}\frac{{\vec K}_{\bp i}\cdot{\vec K}_{\bq j}}{\xi_{\bp i}-\xi_{\bq j}}-\frac{2K_{\bp i}^z}{g_{sc}}.
\en
One then finds
\beg\label{L22}
{\vec {\cal L}}^2(u)=2\sum\limits_{\bp i}\frac{{\cal H}_{\bp i}}{u-\xi_{\bp i}}+\sum\limits_{\bp i}\frac{{\vec K}_{\bp i}^2}{(u-\xi_{\bp i})^2}+\frac{4}{g_{sc}^2}.
\en
From the definition of (\ref{calHp}) we can write 
\beg\label{HmcalHp}
{\cal H}=-\tilde{g}_{sc}\sum\limits_{\bp i}2\xi_{\bp i}{\cal H}_{\bp i}+\textrm{const.}
\en
Since ${\cal L}^2(u)$ is conserved by evolution due to (\ref{LaxPoisson}) and all ${\cal H}_{\bp i}$ Poisson commute with ${\cal L}^2(u)$, it implies that  ${\cal H}_{\bp i}$ are integrable. Furthermore, since $H_{m_z=0}$  is given by a linear combination of ${\cal H}_{\bp i}$ (\ref{HmcalHp}), it also commutes with  ${\cal L}^2(u)$ and therefore is integrable. 

\begin{figure}
\includegraphics[scale=0.26,angle=0]{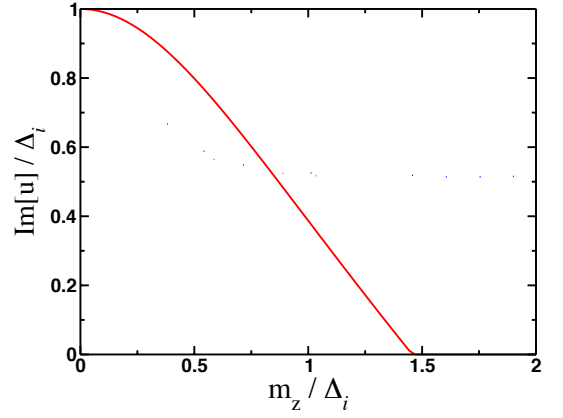}
\caption{(color online) Imaginary part of the root of ${\cal L}^2(u)=0$ as a function of spin-density-wave magnetization 
$m_z$. The parameters are $\Delta_i=0.015\varepsilon_F$ and $\delta_\bk=0$ (no Fermi surface mismatch).}
\label{Fig1}
\end{figure}

\subsection{Lax roots}

To determine the value of the pairing amplitude at long times, we need to compute the imaginary part of the Lax roots governed by the equation
\beg\label{L2eq0}
{\cal L}^2(u)=0.
\en
Using (\ref{Laxfc}) we rewrite (\ref{L2eq0}) as follows,
\beg\label{Eq4Roots}
{\cal L}_z(u)=\pm i{\cal L}_x(u),
\en
where we took into account that initially all $y$ components of the pseudospins are zero, $K_{\bp i}^y=0$. Furthermore, since there is no mismatch between the Fermi surfaces, we have $\xi_{\bp,f}=\frac{p^2}{2}-\mu=-\xi_{\bp,c}=\xi_\bp$. From Eq. (\ref{Epmk}) it follows $E_\bk=\sqrt{\xi_\bk^2+m_z^2+\Delta_0^2}$, where $\Delta_0$ is a superconducting order parameter to be specified below.
Keeping in mind that the $z$-components of ${\vec K}_{\bp\lambda}$ have not been rescaled, for (\ref{Kz}) with
$\xi_f=-\xi_c=\xi$ we have   
\beg\label{Kz4Lax}
\begin{split}
K_c^z(\bk,t=0)&=\frac{\xi_\bk}{2E_\bk}, \quad 
K_f^z(\bk,t=0)=-\frac{\xi_\bk}{2E_\bk}.
\end{split}
\en
Similarly, for the anomalous components (\ref{Kxy}), we find
\beg\label{Kxy4Lax}
\begin{split}
K_c^x(\bk,t=0)&=a_f\left(-\frac{\Delta_0}{2E_\bk}+\frac{\Delta_0}{2E_\bk^3}m_z^2\right), \\
K_f^x(\bk,t=0)&=a_c\left(+\frac{\Delta_0}{2E_\bk}-\frac{\Delta_0}{2E_\bk^3}m_z^2\right).
\end{split} 
\en
Thus, we observe that
\beg\label{SpinsFinal}
K_{\bp i}^z=\frac{a_{i}\xi_\bp}{2E_\bp}=-\frac{\xi_{\bp i}}{2E_\bp}, \quad K_{\bp i}^x=\frac{\Delta_0}{2E_\bp}\left(1-\frac{m_z^2}{E_\bp^2}\right).
\en

\paragraph{Self-consistency conditions.} For quenches into the purely superconducting state, $m_z=0$, a different equilibrium value of the pairing amplitude $\Delta_0$ is determined by the BCS self-consistency 
condition. For a given value of the superconducting coupling $g_{sc}$, we have
\beg\label{gSC}
\frac{2}{g_{sc}}=\sum\limits_\bp\frac{1}{\sqrt{\xi_\bp^2+\Delta_0^2}}.
\en
As a next step, we introduce the function $g_{sc}'(m_z)$ which formally enters as a ``new" coupling constant. The equation which determines this function reads
\beg\label{gSCp}
\frac{2}{g_{sc}'(m_z)}=\sum\limits_\bp\frac{1}{\sqrt{\xi_\bp^2+m_z^2+\Delta_0^2}}.
\en
By comparing (\ref{gSC}) with (\ref{gSCp}) we see that $g_{sc}'(0)=g_{sc}$, so that the imaginary part of the Lax root $\textrm{Im}[u]=\Delta_0$, as it should be for equilibrium. 

\paragraph{Equation for the Lax roots.} Using expressions (\ref{SpinsFinal}) we now rewrite (\ref{Eq4Roots}) as follows. First, momentum summations are replaced with integrals over $\epsilon=p^2/2-\mu$ according to the formula:
\beg\label{Sum2Int}
\sum\limits_{\bp}F(\xi_\bp)=\nu_F\int\limits_{-\mu}^\infty F(\epsilon)d\epsilon
\en
and $\nu_F$ is the density of states at the Fermi level. For the $z$-component of the Lax vector (\ref{Laxfc}) using (\ref{Sum2Int}) we have
\beg\label{Lzint}
\begin{split}
{\cal L}_z(u)&=-\nu_F\int
\limits_{-\mu}^\infty\sum\limits_{\lambda=\pm}\frac{\lambda\epsilon d\epsilon}{2(u-\lambda \epsilon)E(\epsilon)}-\frac{2}{g_{sc}}=\\&=
-\nu_F\beta-\nu_Fu\int
\limits_{-\mu}^\infty\sum\limits_{\lambda=\pm}\frac{d\epsilon}{2(u-\lambda \epsilon)E(\epsilon)},
\end{split}
\en
where we employed Eq. (\ref{gSCp}) and introduced the parameter $\beta$, which describes for the magnitude of the quench
\beg\label{beta}
\beta=2\nu_F^{-1}\left(\frac{1}{g_{sc}}-\frac{1}{g_{sc}'}\right).
\en
Note since $g_{sc}'>g_{sc}$ parameter $\beta$ always remains positive, $\beta>0$. 
We can now use expression (\ref{Lzint}) to rewrite Eq. (\ref{Eq4Roots}) as 
\beg\label{Eq4LaxRoots}
\begin{split}
&\frac{-\beta}{u\pm i\Delta_0}=\left(1\mp\frac{i\Delta_0m_z^2}{u\pm i\Delta_0}\right)\int
\limits_{-\mu}^\infty\sum\limits_{\lambda=\pm}\frac{d\epsilon}{2(u-\lambda \epsilon)E(\epsilon)}.
\end{split}
\en
As we can see from analyzing this equation for $m_z=0$, there will be only one complex root $u=\pm i\Delta_0$. For nonzero $m_z$ we therefore need to find all the complex roots of this equation. The imaginary parts of these roots will determine the value of the superfluid order parameter at long times, see Figs.~\ref{Fig1} and \ref{Fig2}.

\begin{figure}
\includegraphics[scale=0.26,angle=0]{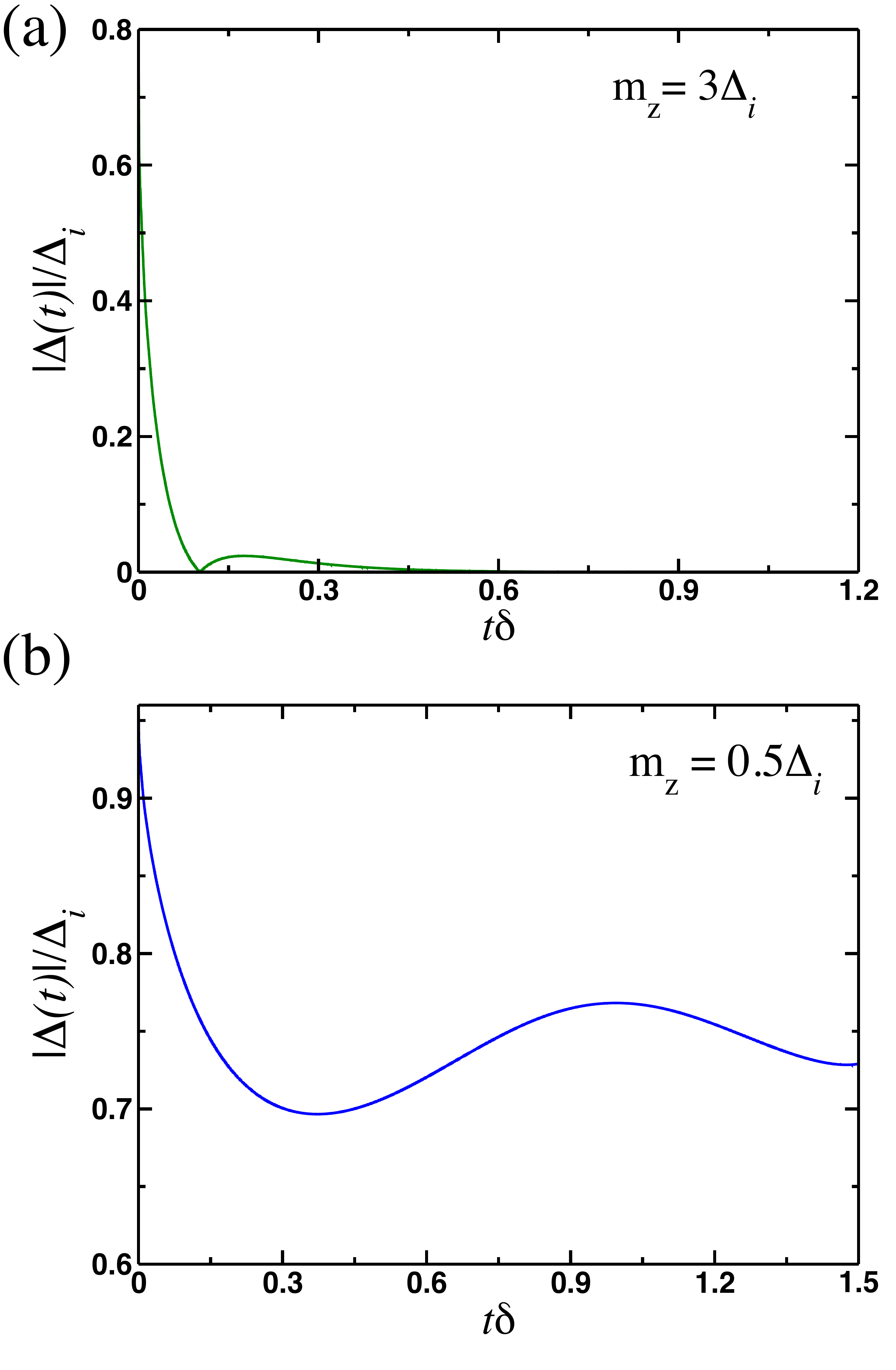}
\caption{(color online) Quenched dynamics of the $s^{\pm}$ pairing amplitude as a function of time ($\delta$ is a level spacing) for initial conditions with a nonzero spin-density-wave magnetization $m_z=3\Delta_i$ (top panel) and $m_z=0.5\Delta_i$ (bottom panel).}
\label{Fig2}
\end{figure}

\section{Discussion and perspectives}\label{Sec-Discussion}  

In this work we have described nonequilibrium kinetics of order parameters in the context of multicomponent superconductors with the emphasis on the iron-pnictide systems. We have found, that out of equilibrium, the coupling between competing superconducting and magnetic orders occurs not only by virtue of self-consistency conditions but also dynamically. This becomes essentially transparent in the pseudospin representation of equations of motion for the Green's function. In particular, as can be seen directly from Eq.~(\ref{EOMKcf}), precession of the superconducting pairing amplitude is strongly affected by the dynamics of magnetic order, which then itself back acts on the $m(t)$. 

Insofar as our analysis is limited to the collisionless regime at time scales satisfying     
\beg
\tau_\Delta\ll t\ll \tau_{in},
\en
where $\tau_\Delta=1/\Delta$, $\tau_{in}$ is the time scale of inelastic scattering processes in the collision-dominated regime. The latter can be found from the golden rule by passing to the Bogoliubov quasiparticle representation where the matrix elements of the transition probabilities in scattering are dressed by the coherence factors. Following the early works of Eliashberg~\cite{Eliashberg} and Galaiko,~\cite{Galaiko} one estimates  
\beg
\tau^{-1}_{in}\sim(g_{sc}\nu_F)^2T^2_c/\varepsilon_F,
\en
which is essentially a Fermi liquid expression for the time scale of electron-electron collisions. It is expected that the power-law decay of the order parameter crosses over to exponential behavior $\propto \exp(-t/\tau_{in})$ once the system enters into the collision-dominated regime of relaxation.  

In the modeling we have adopted the band model of FeSC, which is certainly suitable for clean 122-systems such as isovalently P-doped BaFe$_{2}$(As$_{1-x}$P$_x$)$_2$. It is worth pointing out that in a more general formulation (for example, within the three-band model) additional features may appear, in particular, possibly different branches of collective excitations. It is of clear experimental relevance to revisit the same problem for the disorder model of FeSC,~\cite{FeSc-Dis-Vavilov-1,FeSc-Dis-Vavilov-2} which is more appropriate for the Co-doped case Ba(Fe$_{1-x}$Co$_x$)$_2$As$_2$. However, physically perhaps the most interesting question is to study the nonequilibrium dynamics near the quantum critical point, namely, near the end point of SDW order under a superconducting dome. Such quantum criticality was revealed from the measurements of the London penetration depth~\cite{Matsuda,Auslaender} and attributed to the fluctuations of SDW order at the onset of the transition into the coexistence phase.~\cite{LVKC} The dynamics of magnetization near such a quantum critical point has been recently addressed in the framework of time-dependent Ginzburg-Landau theory,~\cite{Moor-TDGL} however, the nonadiabatic regime still needs to be systematically investigated. In general, post quench prethermalization at a quantum critical point may exhibit nontrivial dynamical scaling.~\cite{Orth} The case of iron pnictides is very specific since the magnetic quantum critical point is surrounded by a superconducting state with gapped quasiparticles, and consequently scaling of the response functions may be governed by entirely different dynamical exponents. Finally, one should seriously look at the role of degrees of freedom associated with the Ising nematic order parameter~\cite{Fernandes-Nematic} that was left behind in our picture. All these questions will pave the way for future research in this field.   


\section*{Acknowledgments} 

We would like to thank A.~Chubukov, R.~Fernandes, P.~Orth, and M.~Schutt for fruitful discussions. 
The work of M.K. was supported by the University of Iowa and Hebrew University of Jerusalem.
The work of A.L. was supported by NSF Grant No. DMR-1401908.

\end{document}